\documentclass{aa}
\usepackage{amssymb}
\usepackage{graphics}
\newcommand{\kmprs}  {\mbox{\rm\,km\,s$^{-1}$}}
\newcommand{\feh} {\mbox{\rm [Fe/H]}}
\newcommand{\teff}  {\mbox{$T_{\rm eff}$}}
\newcommand{\logg}  {\mbox{{\rm log}$g$}}
\newcommand{\Feone} {\ion{Fe}{i}}
\newcommand{\Fetwo} {\ion{Fe}{ii}}
\newcommand{\Sctwo} {\ion{Sc}{ii}}
\newcommand{\Mnone} {\ion{Mn}{i}}

\newcommand{\Vlsr} {\mbox{$V^{'}$}}
\newcommand{\ffe}  {\mbox{${\rm [\frac{Fe}{H}]}$}}
\newcommand{\scfe}  {\mbox{${\rm [\frac{Sc}{Fe}]}$}}
\newcommand{\mnfe} {\mbox{${\rm [\frac{Mn}{Fe}]}$}}

\begin{document}
\thesaurus{07(02.14.1; 08.01.1; 08.12.1; 10.05.1; 10.08.1; 10.19.1)}
\title{Sc and Mn abundances in disk and metal-rich halo stars
\thanks{Based on observations carried out at the European
Southern Observatory, La Silla, Chile, and Beijing Astronomical Observatory,
Xinglong, China}}

\author{P.E.\,Nissen \inst{1} \and Y.Q.\,Chen\inst{1} \inst{2}
  \inst{3} \and W.J.\,Schuster \inst{4} \and G.\,Zhao \inst{3}} 
\offprints{P.E. Nissen}
\institute{Institute of Physics and Astronomy, University of Aarhus, DK--8000
Aarhus C, Denmark
\and  Department of Astronomy, Beijing Normal University, Beijing 100875, China 
\and Beijing Astronomical Observatory, Chinese Academy of Sciences, Beijing 100012, China
\and Observatorio Astronomico National, UNAM, Apartado Postal 877, 22800 Ensenada, B.C., M\'exico}

\date{Received 7 October 1999 / Accepted 8 November 1999}
\maketitle

\begin{abstract}
Sc and Mn abundances are determined for 119 F and G main-sequence stars
with  $-1.4 < \feh < +0.1$, representing stars from the thin disk,
the thick disk and the halo. The results indicate
that Sc behaves like an $\alpha$ element, showing a decreasing
[Sc/Fe] with increasing metallicity in disk stars and a dual pattern in the
kinematically selected halo stars.
In contrast, Mn shows an increase from [Mn/Fe]
$\simeq -0.5$ at $\feh = -1.4$ to zero at solar metallicity. 
There appears to be a discontinuity or sharp increase of [Mn/Fe] at
$\feh \simeq -0.7$ corresponding to the transition between the thick and the
thin disk. It is discussed if supernovae of Type Ia are a major source
of Mn in the Galactic disk or if the trend of [Mn/Fe] vs. [Fe/H] can be
explained by nucleosynthesis in Type II supernovae with
a strong metallicity dependence of the yield.

\keywords{ Nuclear reactions, nucleosynthesis, abundances -- 
Stars: abundances -- Stars: late-type -- 
Galaxy: evolution -- Galaxy: halo -- Galaxy: solar neighbourhood}
\end{abstract}

\section{Introduction}
Scandium and manganese abundances in long-lived F and G stars are of
high interest not only because
they help us to understand the chemical evolution of the Galaxy, but also
because they provide some special constraints on nucleosynthesis
theory.

It has been long known that $\alpha$ elements like O, Mg, Si, and Ca
are mostly
produced by Type II supernovae, while some iron-peak elements have 
significant contributions from Type Ia supernovae. But we have no clear idea
how Sc, as an element intermediate between $\alpha$ elements
and iron-peak elements in the periodic table, is formed.
Sc abundances are available only for a few disk stars, and the two most
extensive works
on halo stars do not give consistent results.
Gratton \& Sneden (\cite{Gratton91}) find a solar Sc/Fe ratio in metal-poor
dwarfs and giants, while Sc is found to be overabundant by Zhao \& Magain 
(\cite{Zhao90}) with [Sc/Fe] $\simeq +0.3$ in metal-poor dwarfs. Clearly,
more detailed abundance information will be useful to reveal the synthesis
history of Sc in the Galaxy.

Mn is known to be quite underabundant with respect to Fe
in metal-poor stars (Gratton \cite{Gratton89}, McWilliam et
al. \cite{McWilliam95}; Ryan et al. \cite{Ryan96}), but the 
pattern of [Mn/Fe] vs. [Fe/H] is not known in great detail.
It is still an open question if Mn is produced mainly in massive stars
as advocated by Timmes et al. (\cite{Timmes95}) or if Type Ia SNe
make a significant contribution at higher metallicities.
Furthermore, the pattern of [Mn/Fe] in disk and metal-rich halo stars
is needed for comparison with recent observations of Mn abundances in
damped Lyman $\alpha$ systems
(Pettini et al. \cite{Pettini99a}; \cite{Pettini99b}).

The reason that the Sc and Mn abundance patterns are not well established
may be related to the significant hyperfine structure (HFS) of
their lines. Data on the HFS of several Sc and Mn lines suitable for
abundance determinations of solar-type dwarfs is, however, available. 
The lack of a consistent study on both elements for disk
and metal-rich halo stars therefore inspired us to carry out a high
precision analysis for a large sample of main-sequence stars selected to have
$5300 < \teff < 6500$~K, $4.0 < \logg < 4.6$, and $-1.4 < \feh < +0.1$.

\section{Observations}
The observational data are taken from two sources:
Chen et al. (\cite{Chen99}, hereafter Chen99) and Nissen \& Schuster
(\cite{Nissen97}, hereafter NS97). The first sample (disk stars) was
observed at Beijing
Astronomical Observatory (Xinglong Station, China)  with the Coud\'{e} Echelle
Spectrograph and a $1024\times 1024$ Tek CCD attached to the 2.16m
telescope, giving a resolution of the order of
40\,000. The second sample (kinematically selected halo stars and metal-poor
disk stars) was observed with the ESO NTT EMMI spectrograph and a $2048 \times
2048$ SITe CCD detector at a higher resolution ($R = 60\,000$). The exposure
times were chosen in order to obtain a signal-to-noise ratio of
above 150 in both samples. 

\begin{figure}
\resizebox{\hsize}{!}{\includegraphics{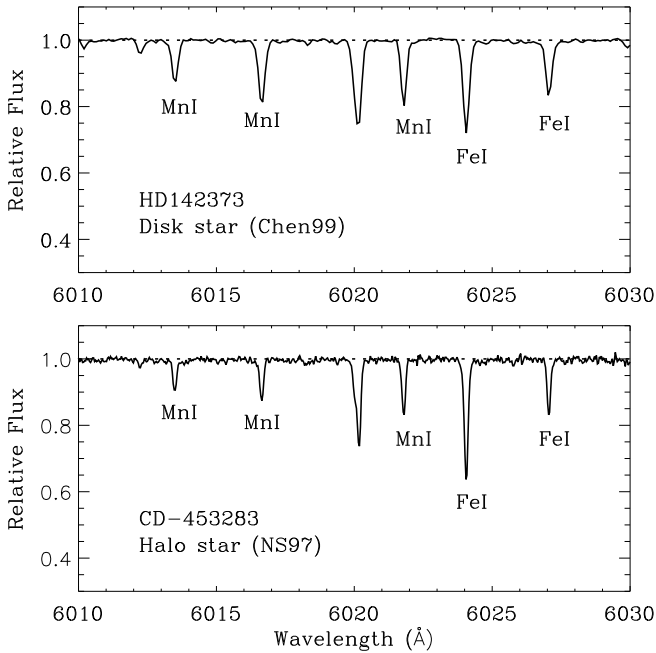}}
\caption{Examples of spectra for one disk star \object{HD\,142373} 
($\teff = 5920 $~K, $\logg = 4.27$, $\feh = -0.39$) obtained with the 2.16m
telescope at Xinglong Station and one halo star \object{CD$-$45\,3283} 
($\teff = 5650 $~K, $\logg = 4.50$, $\feh = -0.84$) observed at the
ESO NTT.}
\label{fig:spec}
\end{figure}

The spectra were reduced with standard MIDAS (Chen99) or IRAF
(NS97) routines for background correction, flatfielding,
extraction of echelle orders, wavelength calibration,
continuum fitting
and measurement of equivalent widths (see Chen99 and NS97 for details).
Figure~\ref{fig:spec} shows
the spectra for one disk star in Chen99 and one halo star in
NS97 around the \Mnone\ triplet at 6020 \AA.

\section{Analysis}
\subsection{Abundance calculation}
In Chen99, the  effective temperature was derived from the
Str\"omgren $b-y$ color index
using the recent calibration by the infrared-flux method (Alonso 
et al. \cite{Alonso96}). For consistency, stellar
temperatures in the NS97 sample (derived from the excitation balance of
\Feone\ lines) were redetermined using the Alonso et
al. calibration of $b-y$. The new \teff 's are on the 
average 60~K lower than those of NS97 but the rms scatter between
the two sets of values is only $\pm 45$~K.

Surface gravities in Chen99 were based on Hipparcos parallaxes
(ESA \cite{ESA97}). For the majority of stars in NS97 accurate
parallaxes are, however, not available,
and surface gravities were therefore determined by requiring
that \Feone\ and  \Fetwo\ lines provide the same iron abundance. As
shown by Chen99, this leads to gravities consistent with those
derived from Hipparcos parallaxes.

Microturbulence velocities were estimated by requesting that 
the iron abundance derived from \Feone\ lines with $\chi > 4$~eV
should not depend on equivalent width.
As shown in Table~\ref{tb:abuerr}, the typical error of the microturbulence
parameter, $\pm 0.3$\kmprs , has no significant influence on 
the derived values of [Sc/Fe] and [Mn/Fe].

The oscillator strengths for two \Sctwo\ lines ($\lambda$5657 and
$\lambda$6604) and two \Mnone\ lines ($\lambda$6013 and $\lambda$6021)
were taken from  Lawler \& Dakin
(\cite{Lawler89}) and  Booth et al. (\cite{Booth84}),
respectively, while differential values for
another three  \Sctwo\ lines  and one \Mnone\ line ($\lambda$6016) were estimated
from an ``inverse'' analysis of 10 well observed
``standard'' stars from Chen99 and all stars from NS97. Hence,
these lines are forced to give the same average abundances of Sc and Mn
as the lines with known $gf$ values for the aforementioned group of stars,
but the inclusion of these lines improves the precision of differential
abundances. We note, that in the spectrum of the Sun,
the \Mnone\ line at 6016~\AA\ appears stronger relative to the two other lines.
Hence, this line may be blended by a weak line in metal-rich stars, but
exclusion of $\lambda$6016 in the derivation of Mn abundances does not
change the derived trend of [Mn/Fe] significantly.

The damping constant corresponding to collisional broadening due
to Van der Waals interaction with hydrogen and helium atoms was
calculated in the Uns\"old (\cite{Unsold55}) approximation with an 
enhancement factor of $E_{\gamma} = 2.5$ for both elements.
The effect of changing the enhancement factor will be discussed in
Sect. 3.3.

The abundance analysis is based on a grid of flux constant,
homogeneous, LTE model atmospheres  computed using the Uppsala
MARCS code with  updated continuous opacities (Asplund et
al. \cite{Asplund97}). Abundance calculations in the LTE approximation
with HFS included
were made using the Uppsala SPECTRUM synthesis
program by forcing the theoretical equivalent widths, derived from
the model, to match the observed ones.

Equivalent widths in the solar flux spectrum were measured from a
Xinglong spectrum of the Moon and analyzed in the same way as the
stellar equivalent widths using the same grid of models 
to interpolate to the atmospheric parameters of the Sun. Hence,
[Sc/Fe] and [Mn/Fe] are derived in a strictly differential way with
respect to the Sun,
thereby minimizing many error sources. In particular, we emphasize
that possible systematic errors in the gf-values do not affect
the relative abundance ratios with respect to the Sun.

Table~\ref{tb:abuew99} and Table~\ref{tb:abuew97} present the
derived abundances, together with the atmospheric parameters
and equivalent widths,
for stars in Chen99 and NS97, respectively.

\subsection{Hyperfine structure effect}      
For all Sc and Mn lines used in our analysis,  it was investigated how
much the HFS affects the derived abundances.
The energy splitting and the relative intensities required in the
HFS calculation for Sc and
Mn are taken from Steffen (\cite{Steffen85}). His data is mostly based on
Biehl (\cite{Biehl76}), but he makes a small adjustment to the wavelength
shift and the relative intensities, leading to fewer components
than Biehl's data. The log$gf$ values for the individual
components (see Table~\ref{tb:hfsdata}) are
then calculated from the relative strengths based on a given
total $gf$ value.

The results show that HFS has a small influence on the two weak
\Sctwo\ lines ($\lambda$6245 and $\lambda$6604) with a maximum of 0.07
dex at $EW
\sim$ 50 m\AA, but the effect is very significant for the stronger
lines ($\lambda$5526 and $\lambda$5657), reaching 0.1-0.3 dex for
the equivalent 
width range of 30-100 m\AA.  
The \Sctwo\ 5239\AA\ line (available in the NS97 spectra) is subject
to a maximum HFS effect of about 0.1 dex for $EW \sim$ 50 m\AA.

For the \Mnone\ lines, the abundance deviation between the
derivations with and without
HFS reaches a maximum of 0.2 dex for the 6013\AA\ line and 0.10
dex for $\lambda 6016$ and $\lambda 6021$ at $EW \sim$ 100 m\AA. We have
found that the HFS data of Booth et
al. (\cite{Booth83}) gives slightly lower Mn abundances than 
Steffen's (\cite{Steffen85}) data; the deviation increases with line
strength, reaching 0.1 dex for the
6013\AA\ line and 0.04 dex for $\lambda 6016$ and $\lambda 6021$ at
$EW \sim$ 100 m\AA. The reason for the discrepancy is
unclear. But the original HFS data for $\lambda$6013 from Biehl (\cite{Biehl76})
gives exactly the same abundance as that based on Steffen's
(\cite{Steffen85}) HFS data.
 
The final abundances are the mean values of the abundances derived
from the available lines with
HFS according to Steffen (\cite{Steffen85}) taken into account.

\begin{table}
\caption[ ]{The hyperfine structure data for Sc and Mn. $\chi$ is the excitation
potential of the lower energy level}
\label{tb:hfsdata}
\begin{tabular}{rrr|rrr}
\noalign{\smallskip}
\hline
\noalign{\smallskip}
  $\lambda$[\AA ] & $\chi$[eV] & $\log gf$ &  $\lambda$[\AA ] &$\chi$[eV] & $\log gf$ \\
\noalign{\smallskip}
\hline
\Sctwo & & & \Sctwo & & \\
5239.779 & 1.45  &$-$1.468 & 5526.777    & 1.77  & $-$0.858 \\
5239.812 & 1.45  &$-$1.375 & 5526.810    & 1.77  & $-$0.765 \\
5239.845 & 1.45  &$-$1.583 & 5526.843    & 1.77  & $-$0.973  \\ 
5239.864 & 1.45  &$-$1.472 & 5526.862    & 1.77  & $-$0.862 \\
\hline
\noalign{\smallskip}
\Sctwo & & & \Sctwo & & \\
5657.836    & 1.51  & $-$1.205 &6245.576   &  1.51  & $-$1.736  \\
5657.869    & 1.51  & $-$1.112 &6245.609   &  1.51  & $-$1.643  \\
5657.902    & 1.51  & $-$1.320 &6245.642   &  1.51  & $-$1.851  \\
5657.921    & 1.51  & $-$1.209 &6245.661   &  1.51  & $-$1.740 \\
\noalign{\smallskip}
\hline
\Sctwo & & & \Mnone & & \\
     6604.556    & 1.36  & $-$1.911 &6013.537    & 3.07  & $-$1.365  \\
     6604.589    & 1.36  & $-$1.818 &6013.519    & 3.07  & $-$0.787   \\
     6604.622    & 1.36  & $-$2.026 &6013.501    & 3.07  & $-$1.108   \\
     6604.641    & 1.36  & $-$1.915 &6013.486    & 3.07  & $-$0.977\\
                 &       &        &6013.474    & 3.07  & $-$0.767 \\
\noalign{\smallskip}
\hline
\Mnone & & & \Mnone & & \\
6016.665  &    3.07  &  $-$0.702 &6021.814   &   3.07  &  $-$0.477\\
6016.650  &    3.07  &  $-$1.345 &6021.806   &   3.07  &  $-$0.612\\
6016.638  &    3.07  &  $-$0.683 &6021.797   &   3.07  &  $-$0.395 \\
6016.623  &    3.07  &  $-$0.588 &6021.780   &   3.07  &  $-$1.225 \\  
6016.604  &    3.07  &  $-$1.071 &6021.764   &   3.07  &  $-$1.374 \\
\noalign{\smallskip}
\hline
\end{tabular}
\end{table}

\subsection{Abundance uncertainties}
Abundance errors are mainly due to
random errors in the equivalent widths, the
uncertainty of the stellar atmospheric parameters, and a possible error
in the enhancement factor of the damping constant. 
An estimate of the first contribution is obtained by dividing the
rms scatter of abundances, determined from the individual lines,
by $\sqrt{N}$, where $N$ is the number of lines. The second kind of
uncertainty is estimated by a change of 70 K in $\teff$, 0.1 dex
in $\logg$, 0.1 dex
in \feh\ and 0.3\kmprs\ in  microturbulence, typical standard deviations
for these atmospheric parameters as estimated in Chen99. Finally, the abundance
change caused by a variation of $E_{\gamma}$ by 50\% was calculated.

Table~\ref{tb:abuerr} presents the errors in the relative abundances
for a typical disk star and one of the halo stars. 
Note, that we have compared the
Sc abundance derived from \Sctwo\ lines with the Fe abundance derived
from \Fetwo\ lines and the Mn abundance derived from \Mnone\ lines with the
Fe abundance derived from \Feone\ lines. Hence, the ratios are derived from
lines having a similar dependence on \teff\ and \logg , which explains the
the rather small effect of the uncertainty in the atmospheric parameters.
In the case of Sc the uncertainty of the enhancement factor is not important,
because the \Sctwo\ lines are quite weak and not of too different strenghts
in the solar and the stellar spectra. 
The \Mnone\ lines show, however, a large variation in equivalent width,
from about 100~m\AA\ at solar metallicity to about 10~m\AA\ in the metal-poor
halo stars (due to the underabundance of Mn with respect to Fe). Hence,
the choice of $E_{\gamma}$ has a significant effect on the slope of [Mn/Fe]
vs. [Fe/H]. As seen from Table~\ref{tb:abuerr}, an increase of
$E_{\gamma}$ by 50\% (from 2.5 to 3.75) increases the derived [Mn/Fe]
at $\feh \sim -0.8$ by 0.05 dex, and a decrease of $E_{\gamma}$ from
2.5 to 1.0 (i.e. no enhancement of the Uns\"old approximation for the damping
constant) would decrease [Mn/Fe] of the metal-poor stars by about 0.10
dex. Unfortunately, there are no reliable theoretical calculations of
the damping constant for the \Mnone\ lines, nor any empirical estimates of
$E_{\gamma}$ from e.g. the solar spectrum. 

In addition to the abundance errors estimated in Table 2, possible
non-LTE effects should be considered. As the Sc abundance is determined
from \Sctwo\ lines and compared to Fe from \Fetwo\ lines,
the non-LTE effects on [Sc/Fe] should be small, because the abundances of
both elements are based on lines from the dominating ionization stage.
The Mn abundances (derived from \Mnone\ lines) may, however, be subject
to non-LTE effects due to overionization of \Mnone\ caused by the UV
radiation field, especially in the metal-poor stars. According to our
knowledge there is no non-LTE calculations of Mn for F and G stars, but noting
that the ionization potential of \Mnone\ (7.43 eV) is similar to that
of \Feone\ (7.87 eV) one might expect that the ratio Mn/Fe is not
significantly changed by non-LTE effects, when the abundances of both
elements are determined from neutral lines. Clearly, this should be checked
by detailed computations.

\begin{figure*}
\resizebox{17cm}{!}{\includegraphics{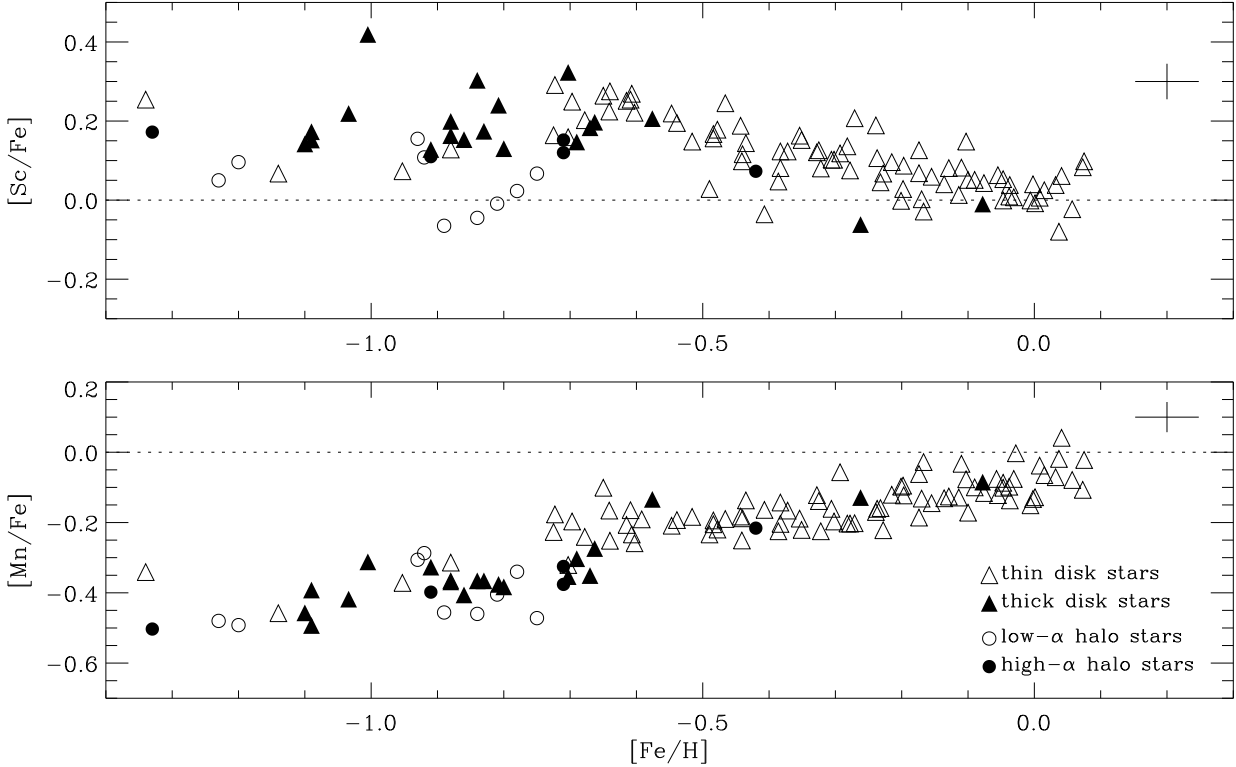}} \hfill
\caption{Abundance patterns for Sc and Mn in four groups of stars.} 
\label{fig:ScMn}
\end{figure*}

\begin{table}
\caption[]{Abundance errors for \object{HD\,142373} and \object{CD$-$45\,3283}.
The atmospheric parameters of the stars are given in connection with 
Fig.~\ref{fig:spec}}
\label{tb:abuerr}
\begin{tabular}{lrrrr}
\noalign{\smallskip}
\hline
\noalign{\smallskip}
 & \multicolumn{2}{c}{HD\,142373} & \multicolumn{2}{c}{CD$-$45\,3283}\\
 & $\Delta \scfe$ & $\Delta \mnfe$  & $\Delta \scfe$ & $\Delta \mnfe$ \\
\noalign{\smallskip}
\hline
$\frac{\sigma_{EW}}{\sqrt{N}}$&      0.033   &0.037    &   0.045 &0.042  \\
$\Delta \teff = +70$K &                0.023   &0.002    &   0.011 &0.017  \\
$\Delta \log g = +0.1$ &            $-$0.001   &$-$0.002 &$-$0.006 &0.003  \\
$\Delta \feh = +0.1$  &                0.004   &0.001    &   0.004 &0.003  \\
$\Delta \xi = +0.3$ &               $-$0.011   &0.008    &   0.008 &0.018  \\
$\Delta E_{\gamma} = +50\%$ &          0.002   &0.040    &   0.006 &0.052  \\
\noalign{\smallskip}
\hline
\end{tabular}
\end{table}

\section{Discussion and conclusions}
Stars in the metallicity range $-1.4 < \feh < +0.1$ include a mixture
of different
populations. In our sample, three groups can be separated according to
their kinematics: thin disk stars ($\Vlsr > -50\kmprs$),
thick disk stars ($-120 < \Vlsr < -50\kmprs$) and  halo stars 
($\Vlsr < -175\kmprs$). The values of \Vlsr (the velocity component in
the direction of Galactic rotation with respect to the Local Standard of
Rest) are taken from Chen99 and
NS97. As discussed in Chen99 a rotational lag of about 50\kmprs\ corresponds 
rather well to
the transition between thick disk stars ($\sigma (W^{'}) \sim 40$\kmprs\ and
ages $>10$~Gyr) and thin disk stars ($\sigma (W^{'}) \sim 20$\kmprs\ and
ages $<10$~Gyr) although a clean separation between the two populations
is not possible. The thick disk stars may also be contaminated by a 
few halo stars, whereas the group with $\Vlsr < -175\kmprs$ consists
of genuine halo stars according to the Galactic orbits computed in NS97.
Furthermore, the halo stars can be split up into
``low-$\alpha$'' and  ``high-$\alpha$''  stars according to the
abundances of $\alpha$ elements as described by NS97.
The low-[$\alpha$/Fe] halo stars may belong to the
``high-halo'', resulting from a merging process of the Galaxy with 
satellite components (NS97), or they may come from low-density regions in the
outer halo where the chemical evolution proceeded 
slowly allowing incorporation of iron from Type Ia SNe at a lower
metallicity than in the inner halo and the thick disk
(NS97; Gilmore \& Wyse \cite{Gilmore98}).

Figure~\ref{fig:ScMn} shows the run of [Sc/Fe] and [Mn/Fe]
vs. [Fe/H] with the four groups of stars marked by different
symbols.

\subsection{Scandium}

As seen from Fig.~\ref{fig:ScMn}, [Sc/Fe] declines from an
overabundance ($\sim$ 0.2) at $\feh = -1.4$ to zero
at solar metallicity even though there are some exceptions. 
Hence, Sc seems to follow the even-Z $\alpha$ elements like Si, Ca and Ti
(see Edvardsson et al. \cite{Edvardsson93} and Chen99). 
In particular, Sc keeps a  near-constant overabundance
for $\feh <-0.6$ like the $\alpha$ elements, except for the
group of low-[$\alpha$/Fe] halo stars, which tend to have low
values of [Sc/Fe], just as expected if Sc belongs to the $\alpha$-element
family.

Theoretically, Samland's (\cite{Samland98}) model
reproduces both the overabundance of about 0.2 dex in [Sc/Fe] at
$\feh \sim -1.0$ and the decreasing relation with increasing
metallicity by suggesting that  Sc is mostly made by
high mass stars. Observationally,
our result supports the high values of [Sc/Fe] for metal-poor dwarfs by
Zhao \& Magain (\cite{Zhao90}) 
instead of the solar ratio found by Gratton \& Sneden
(\cite{Gratton91}) for metal-poor dwarfs and giants.
The reason for this discrepancy
is unclear, but we note that Gratton \& Sneden 
adopt solar $gf$ values based on the empirically-derived  solar model
by Holweger \& M\"uller (\cite{Holweger74}). 
As stated in their paper, the use of 
a theoretical model for the Sun (like for the stars) would increase
the derived [Sc/Fe] values by 0.06 dex. Furthermore, we note that
Hartmann \& Gehren (\cite{Hartmann88}) derive an average value
[Sc/Fe] = +0.11 for 16 metal-poor subdwarfs.

\subsection{Manganese}
Figure~\ref{fig:ScMn} shows that [Mn/Fe] increases from a very significant
underabundance at [Fe/H] $\sim -$ 1.4 to a solar ratio 
at $\feh \simeq 0.0$. The underabundance of $\sim -0.5$ in the
metal-poor stars is consistent with the recent study of Mn in damped
Lyman-$\alpha$ systems by Pettini et al. 
(\cite{Pettini99a}; \cite{Pettini99b}), and with
the results of Gratton (\cite{Gratton89}), who found [Mn/Fe] $\sim -0.4$
in 11 giants and dwarfs with $-2.5 < \feh < -1.0$ and an increasing
trend of [Mn/Fe] with [Fe/H] for 14 disk stars.

Timmes et al. (\cite{Timmes95}) have shown that nucleosynthesis of Mn
in massive stars with a metallicity dependent yield (due to the lower
neutron excess in metal-poor stars) explains the trend of [Mn/Fe]
rather well, and they argue that Type Ia SNe are not a major
contributor to the synthesis of Mn. As seen from Fig.~\ref{fig:ScMn} 
there is, however, a rather sharp increase of [Mn/Fe] around 
$\feh \sim -0.7$ or even a discontinuity in [Mn/Fe] between the thick
disk and halo stars on one side and the thin disk stars on the other
side. In fact, the trend of [Mn/Fe] appears to mirror that of the 
$\alpha$ elements, especially that of O (Edvardsson et al.
\cite{Edvardsson93}; Chen99). This suggests that Type Ia SNe 
give a large contribution to the production of Mn in the Galactic disk
in agreement with the chemical evolution model of Samland (\cite{Samland98}),
who finds that 75\% of the Galactic Mn is produced in SNe of Type Ia.
On the other hand, we note that the ``low-$\alpha$'' halo stars do not
stand out from the thick disk stars in the [Mn/Fe] -- [Fe/H] diagram.
If Type Ia SNe make a large contribution to the production of Mn
one might expect the ``low-$\alpha$'' stars to have higher [Mn/Fe]
values than the thick disk stars, because the ``low-$\alpha$'' stars
were explained by assuming that they were formed
from interstellar gas enriched at lower than usual \feh\ values
with the products of Type Ia supernovae.
However, the position of the
"low-$\alpha$" stars in Fig. 2 may be due to cancelling effects; the Type
Ia SNe produce no O and fewer $\alpha$-elements than Type II SNe,
and the neutron excess necessary
for Mn production may depend more on the overall metallicity than on
just [Fe/H].

We conclude that the nucleosynthesis of Mn may be modulated in a
complicated way.
Moreover, it is hard to understand why a strong underabundance like that of 
Mn is not present for the other odd-Z iron-peak elements, V and Co.
According to the results of Gratton \& Sneden (\cite{Gratton91}),
it seems that V and Co vary in lockstep with Fe down to metallicities
of $\feh \simeq -2.5$, perhaps with a slight underabundance of Co.
Furthermore, Mn and Co show  very different behaviours below 
$\feh \simeq -2.5$ (McWilliam et al. \cite{McWilliam95}; Ryan et al. \cite{Ryan96})
with [Mn/Fe] decreasing strongly and [Co/Fe] increasing. Hence,
the term ``iron-peak'' elements does not indicate the products of a single
nuclear reaction. These elements may have different origins. Further
study of Mn and of other ``iron-peak'' elements is desirable to understand
their synthesis.

\section*{Acknowledgements}
This research was supported by the Danish Research Academy and
the Chinese Academy of Science, and by CONACyT and DGAPA, UNAM in M\'exico.

\begin{table*}
\caption{Atmospheric parameters, relative abundance ratios, and 
equivalent widths in m\AA\ for 4 \Sctwo\ and 3 \Mnone\ lines for stars from
Chen99.}
\label{tb:abuew99}
\begin{tabular}{lrrrrrrrrrrrrr}
\noalign{\smallskip}
\hline
\noalign{\smallskip}
Star (HD)& \teff & \logg & \ffe & \scfe & \mnfe      
&  $\lambda$5526&  $\lambda$5657&  $\lambda$6245&  $\lambda$6604&  $\lambda$6013&  $\lambda$6016&  $\lambda$6021 \\
\noalign{\smallskip}
\hline
   SUN &5780& 4.44&  0.00&  0.00&  0.00& 80.6       & 71.6 & 38.4 & 39.3 & 92.7 &100.5 & 99.0\\
   400 &6122& 4.13& $-$0.23&  0.05& $-$0.16&            & 71.0 & 30.8 & 34.8 & 45.7 & 65.4 & 53.9\\
  3454 &6056& 4.29& $-$0.59&      & $-$0.19&            & 55.5 & 22.2 &      &      & 32.8 & 40.6\\
  5750 &6223& 4.21& $-$0.44&  0.12& $-$0.19&            & 54.2 & 25.0 &      &      & 40.7 & 40.3\\
  6834 &6295& 4.12& $-$0.73&  0.29& $-$0.18&            & 48.3 & 22.6 &      & 13.3 & 25.5 &     \\
  6840 &5860& 4.03& $-$0.45&  0.19& $-$0.18&            & 71.4 & 39.8 & 31.4 & 45.4 & 53.2 & 60.4\\
 10307 &5776& 4.13& $-$0.05&      & $-$0.08&            & 77.1 & 43.3 & 41.5 & 84.1 & 98.5 & 92.4\\
 11007 &6027& 4.20& $-$0.16&  0.06& $-$0.14&            & 61.8 & 42.2 & 34.1 & 56.6 & 65.9 &     \\
 11592 &6232& 4.18& $-$0.41& $-$0.04& $-$0.16&            & 60.0 & 13.9 &      & 34.3 & 37.9 &     \\
 19373A&5867& 4.01&  0.03&  0.04& $-$0.07&            & 85.9 & 49.7 & 51.7 & 86.5 &      & 95.9\\
 22484 &5915& 4.03& $-$0.13&  0.08& $-$0.13&            & 81.0 & 42.7 & 44.1 & 67.4 & 80.5 & 80.6\\
 24421 &5986& 4.10& $-$0.37&  0.12& $-$0.17&            & 71.4 & 28.5 & 31.2 & 43.5 & 52.5 & 57.9\\
 25173 &5867& 4.07& $-$0.62&  0.25& $-$0.17&            & 64.4 & 30.2 & 28.1 & 28.8 & 44.4 & 49.9\\
 25457 &6162& 4.28& $-$0.11&  0.05& $-$0.17&            &      & 29.9 & 39.5 & 62.3 & 70.8 & 71.4\\
 25998 &6147& 4.35& $-$0.11&  0.15& $-$0.08&            &      & 36.1 & 44.9 &      & 89.4 & 87.9\\
 33632A&5962& 4.30& $-$0.23&  0.07& $-$0.22&            & 64.4 & 30.4 & 29.3 & 43.5 & 64.7 & 64.7\\
 34411A&5773& 4.02&  0.01&  0.01& $-$0.04&            & 83.8 & 45.3 & 46.2 & 89.3 &      & 99.0\\
 35296A&6015& 4.24& $-$0.14&  0.04& $-$0.13&       81.0 &      &      & 34.0 &      & 81.5 & 72.6\\
 39587 &5805& 4.29& $-$0.18&  0.13& $-$0.06&       74.9 & 76.1 & 32.1 & 33.2 & 87.4 & 87.6 & 87.4\\
 39833 &5767& 4.06&  0.04& $-$0.08& $-$0.02&       82.8 & 77.3 & 43.2 & 43.0 & 95.1 &      &     \\
 41640 &6004& 4.37& $-$0.62&  0.25& $-$0.21&            & 56.3 & 23.2 & 20.9 & 30.7 & 31.8 & 38.1\\
 43947 &5859& 4.23& $-$0.33&  0.13& $-$0.14&       73.6 &      & 30.7 & 26.8 & 60.3 & 62.1 & 65.5\\
 46317 &6216& 4.29& $-$0.24&  0.19& $-$0.17&       93.0 & 75.9 & 32.3 & 30.0 & 43.7 & 51.8 & 53.5\\
 49732 &6260& 4.15& $-$0.70&  0.25& $-$0.20&            &      & 21.0 & 18.5 & 19.3 &      & 25.2\\
 54717 &6350& 4.26& $-$0.44&  0.14& $-$0.14&       61.2 & 57.6 & 23.5 & 22.6 & 35.0 & 34.5 & 35.4\\
 55575 &5802& 4.36& $-$0.36&  0.16& $-$0.19&       72.5 & 61.2 & 33.1 & 30.0 & 52.7 & 60.3 & 64.8\\
 58551 &6149& 4.22& $-$0.54&  0.19& $-$0.19&       70.6 & 56.8 & 23.5 & 20.3 & 30.1 & 33.2 & 37.8\\
 58855 &6286& 4.31& $-$0.31&  0.10& $-$0.16&            & 63.3 & 26.5 & 24.3 & 38.7 & 44.6 & 48.8\\
 59380 &6280& 4.27& $-$0.17&  0.00& $-$0.13&       74.1 &      & 27.6 & 26.8 & 48.1 & 59.4 & 59.8\\
 59984A&5900& 4.18& $-$0.71&  0.16& $-$0.32&       61.3 & 54.7 & 25.0 & 24.6 &      & 29.6 & 29.8\\
 60319 &5867& 4.24& $-$0.85&  0.30& $-$0.37&       54.8 & 49.7 & 21.0 & 17.3 & 13.7 & 19.0 & 26.8\\
 62301 &5837& 4.23& $-$0.67&  0.20& $-$0.27&       63.2 & 55.5 & 27.2 & 23.2 & 25.7 & 33.9 & 38.7\\
 63333 &6057& 4.23& $-$0.39&  0.12& $-$0.14&       68.0 & 63.7 & 26.1 & 29.1 & 48.5 & 47.7 & 50.4\\
 68146A&6227& 4.16& $-$0.09&  0.05& $-$0.10&       85.8 & 84.6 & 38.6 & 33.0 & 63.0 & 65.5 & 70.4\\
 69897 &6243& 4.28& $-$0.28&  0.14& $-$0.20&       73.3 & 80.0 & 29.1 & 25.4 & 40.6 & 44.7 & 48.7\\
 72945A&6202& 4.18&  0.00&  0.00& $-$0.15&       91.6 & 84.0 & 39.4 & 35.2 &      & 77.2 & 76.7\\
 75332 &6130& 4.32&  0.00& $-$0.01& $-$0.13&       86.5 & 82.0 & 31.6 & 36.8 & 76.8 & 81.8 & 77.7\\
 76349 &6004& 4.21& $-$0.49&  0.17& $-$0.19&       77.5 & 61.8 & 26.7 & 21.2 & 34.0 & 43.8 & 49.3\\
 78418A&5625& 3.98& $-$0.26& $-$0.06& $-$0.13&       64.6 & 68.3 & 33.6 & 34.5 & 73.6 & 76.7 & 74.0\\
 79028 &5874& 4.06& $-$0.05&  0.05& $-$0.09&       92.1 & 90.8 & 46.3 & 48.3 & 84.3 & 90.8 & 87.0\\
 80218 &6092& 4.14& $-$0.28&  0.08& $-$0.20&       78.3 & 64.6 & 37.0 & 27.5 &      & 55.0 & 53.7\\
 89125A&6038& 4.25& $-$0.36&  0.15& $-$0.22&       72.7 & 73.5 & 29.8 & 25.3 &      & 46.7 & 52.8\\
 90839A&6051& 4.36& $-$0.18&  0.07& $-$0.19&       77.1 & 67.5 & 33.7 & 30.4 & 56.4 & 67.4 & 65.2\\
 91889A&6020& 4.15& $-$0.24&  0.11& $-$0.16&       87.1 & 71.9 & 33.8 & 34.6 & 51.6 & 59.6 & 62.7\\
 94280 &6063& 4.10&  0.06& $-$0.02& $-$0.08&       94.3 & 86.8 & 47.3 & 37.5 & 82.0 & 92.7 & 83.9\\
 95128 &5731& 4.16& $-$0.12&  0.08& $-$0.03&            & 77.4 & 42.2 & 40.8 & 94.7 & 95.8 & 90.9\\
 97916 &6445& 4.16& $-$0.94&  0.07& $-$0.37&            &      &  8.1 &      &      &      & 10.4\\
100180A&5866& 4.12& $-$0.11&  0.01& $-$0.13&       88.9 & 76.4 & 36.0 & 34.4 & 72.2 & 84.7 & 82.7\\
100446 &5967& 4.29& $-$0.48&  0.18& $-$0.22&       66.1 & 60.9 & 26.4 & 25.5 & 32.4 & 44.7 &  0.0\\
100563 &6423& 4.31& $-$0.02&      &  0.00&            &      &      &      & 70.9 &      & 75.8\\
101676 &6102& 4.09& $-$0.47&  0.25& $-$0.19&       94.4 & 64.1 & 28.7 & 32.2 & 33.5 & 42.1 & 40.6\\
106516A&6135& 4.34& $-$0.71&  0.32& $-$0.35&       62.1 & 48.9 & 22.3 & 19.9 & 16.0 & 20.3 & 18.8\\
108510 &5929& 4.31& $-$0.06&  0.06& $-$0.12&       90.2 & 78.1 & 38.5 & 34.7 & 78.7 & 86.5 & 84.2\\
109303 &5905& 4.10& $-$0.61&  0.27& $-$0.23&       73.6 & 68.7 & 24.7 & 30.9 &      & 34.2 & 44.6\\
114710A&5877& 4.24& $-$0.05&  0.00& $-$0.10&       81.7 & 73.6 & 37.2 & 34.8 & 81.0 & 87.1 & 86.9\\
\noalign{\smallskip}
\hline
\end{tabular}
\end{table*}

\begin{table*}
\indent
{\bf Table 3.}(continued)~~~~~~~~~~~~~~~~~~~~~~~~~~~~~~~~~~~~~~~~~~~~~~~~~~~~~~~~\\[3mm]
\begin{tabular}{lrrrrrrrrrrrrr}
\noalign{\smallskip}
\hline
\noalign{\smallskip}
Star (HD)& \teff & \logg & \ffe & \scfe & \mnfe 
&  $\lambda$5526&  $\lambda$5657&  $\lambda$6245&  $\lambda$6604&  $\lambda$6013&  $\lambda$6016&  $\lambda$6021 \\
\noalign{\smallskip}
\hline
115383A&5866& 4.03&  0.00&  0.04& $-$0.13&       97.4 & 99.8 & 45.9 & 43.8 &      &      & 96.3\\
118244 &6234& 4.13& $-$0.55&  0.22& $-$0.21&       78.4 & 58.2 & 22.5 & 24.8 & 27.2 & 30.3 & 31.9\\
121560 &6059& 4.35& $-$0.38&  0.08& $-$0.20&       73.8 & 53.5 & 24.5 & 22.1 & 35.8 & 49.5 & 49.2\\
124244 &5853& 4.11&  0.05&  0.06&  0.04&            & 85.2 & 50.0 & 48.9 & 96.9 &      &     \\
128385 &6041& 4.12& $-$0.33&  0.12& $-$0.12&       75.4 & 74.2 & 34.9 & 31.6 & 48.5 & 54.4 & 64.9\\
130948 &5780& 4.18& $-$0.20&  0.03& $-$0.09&       82.9 & 70.4 & 34.8 & 30.1 & 80.0 & 84.5 & 85.4\\
132254 &6231& 4.22&  0.07&  0.08& $-$0.11&       97.3 & 96.4 & 43.7 & 47.3 & 72.4 & 80.0 & 79.8\\
139457 &5941& 4.06& $-$0.52&  0.15& $-$0.18&            & 64.8 & 23.3 & 29.7 &      & 43.0 & 52.2\\
142373 &5920& 4.27& $-$0.39&  0.05& $-$0.22&       71.6 & 66.9 & 31.9 &      & 36.5 & 55.5 & 52.8\\
142860A&6227& 4.18& $-$0.22&  0.10& $-$0.12&            &      &      & 31.4 & 51.4 & 56.3 & 59.5\\
146099A&5941& 4.10& $-$0.61&  0.22& $-$0.26&            & 63.1 & 25.4 &      & 29.0 & 32.9 & 36.9\\
149750 &5792& 4.17&  0.08&  0.10& $-$0.02&            &      & 51.1 &      & 99.2 &      &     \\
154417 &5925& 4.30& $-$0.04&  0.01& $-$0.10&       83.1 & 75.2 & 34.3 & 34.2 & 76.3 & 90.2 & 85.6\\
157347 &5654& 4.36& $-$0.02&  0.01& $-$0.08&            & 72.3 & 38.3 &      & 94.6 &      &     \\
157466 &5935& 4.32& $-$0.44&  0.10& $-$0.25&            & 54.7 & 23.9 &      & 37.3 & 46.5 & 50.1\\
162004B&6059& 4.12& $-$0.08&  0.04& $-$0.12&            & 84.7 & 41.9 & 36.8 & 69.5 &      & 82.2\\
167588 &5894& 4.13& $-$0.33&  0.08& $-$0.22&            & 72.5 & 39.5 & 37.1 & 44.4 & 57.6 & 60.7\\
168009 &5719& 4.08& $-$0.07& $-$0.01& $-$0.09&            & 78.5 & 40.1 & 37.2 & 84.6 &      & 97.1\\
170153A&6034& 4.28& $-$0.65&  0.26& $-$0.10&            & 44.6 & 23.8 & 29.7  &     & 43.6 & 40.8\\
184601 &5830& 4.20& $-$0.81&  0.24& $-$0.38&            & 48.5 & 19.3 & 18.7 & 13.5 & 25.1 & 27.0\\
189340 &5888& 4.26& $-$0.19&  0.09& $-$0.12&            & 67.0 & 33.6 & 40.7 & 68.4 & 80.8 & 83.6\\
191862A&6328& 4.19& $-$0.27&  0.21& $-$0.20&            & 74.4 & 38.3 & 39.9 & 37.9 & 45.2 & 51.7\\
198390 &6339& 4.20& $-$0.31&  0.10& $-$0.20&            & 67.2 &      & 24.4 & 39.4 & 43.8 & 34.1\\
200580 &5829& 4.39& $-$0.58&  0.21& $-$0.13&            & 51.7 & 21.0 & 25.9 & 32.6 & 56.6 & 53.3\\
201891 &5827& 4.43& $-$1.04&  0.22& $-$0.42&            & 29.7 & 18.0 &  8.7 &  9.7 &      & 15.9\\
204306 &5896& 4.09& $-$0.65&  0.28& $-$0.25&            & 58.1 & 31.8 & 27.5 & 24.0 & 34.6 & 40.1\\
204363 &6141& 4.18& $-$0.49&  0.16& $-$0.20&            & 54.5 & 25.6 & 26.5 & 29.4 & 36.0 & 42.2\\
206301 &5682& 3.98& $-$0.04&  0.04& $-$0.14&            & 85.6 & 54.6 & 46.8 & 94.0 &      & 96.5\\
206860 &5798& 4.25& $-$0.20&  0.00& $-$0.10&            & 65.5 &      & 30.6 & 73.6 & 92.9 & 87.7\\
208906A&5929& 4.39& $-$0.73&  0.16& $-$0.23&            & 38.8 & 15.8 & 17.7 & 20.6 & 31.9 & 34.5\\
209942A&6022& 4.25& $-$0.29&  0.12& $-$0.06&            & 67.6 & 36.4 & 28.4 & 54.2 & 78.9 & 67.1\\
210027A&6496& 4.25& $-$0.17& $-$0.03& $-$0.03&            & 60.1 & 24.4 &      & 48.4 & 62.0 & 46.0\\
210752 &5847& 4.33& $-$0.68&  0.20& $-$0.24&            & 45.3 & 16.8 & 27.3 & 30.1 & 33.9 & 41.6\\
212029A&5875& 4.36& $-$1.01&  0.42& $-$0.31&            & 35.9 & 20.1 & 17.6 & 18.9 & 11.6 & 18.1\\
215257 &5976& 4.36& $-$0.65&  0.22& $-$0.17&            & 50.6 &      &      & 23.6 & 49.3 & 34.0\\
219623A&6039& 4.07&  0.02&  0.03& $-$0.06&            & 80.4 & 40.5 & 39.8 & 77.4 & 98.1 & 73.7\\
\noalign{\smallskip}
\hline
\end{tabular}
\end{table*}

\begin{table*}
\caption{Atmospheric parameters, relative abundance ratios, and equivalent
widths in m\AA\ for 5 \Sctwo\ and 3 \Mnone\ lines for stars from NS97}
\label{tb:abuew97}
\begin{tabular}{lrrrrrrrrrrrrrr}
\noalign{\smallskip}
\hline
\noalign{\smallskip}
Star & \teff & \logg & \ffe & \scfe & \mnfe     
&  $\lambda$5239&  $\lambda$5526&  $\lambda$5657&  $\lambda$6245&  $\lambda$6604&  $\lambda$6013&  $\lambda$6016&  $\lambda$6021\\
\noalign{\smallskip}
\hline
BD$-$21\,3420&5858& 4.25& $-$1.09&  0.17& $-$0.49&    18.4 & 37.7 & 28.0 & 10.2 & 11.5 &  8.1 &  9.4 & 10.5\\
CD$-$33\,3337&6022& 3.99& $-$1.34&  0.25& $-$0.34&    17.1 & 37.1 & 26.1 &  7.9 &  8.0 &  4.6 &  6.8 &  8.0\\
CD$-$45\,3283&5650& 4.50& $-$0.84& $-$0.05& $-$0.46&    18.0 & 33.3 & 23.1 &  8.0 & 10.1 & 16.0 & 21.6 & 28.3\\
CD$-$47\,1087&5657& 4.20& $-$0.80&  0.13& $-$0.38&    32.5 & 53.8 & 43.5 & 18.6 &      & 20.2 & 27.0 & 32.8\\
CD$-$57\,1633&5944& 4.22& $-$0.89& $-$0.06& $-$0.45&    17.7 & 36.1 & 24.3 &  7.9 &  8.9 &  9.5 & 13.9 & 18.4\\
CD$-$61\,0282&5772& 4.20& $-$1.23&  0.05& $-$0.48&    11.8 & 25.6 & 19.3 &  6.6 &  7.8 &  6.0 &  7.4 & 11.4\\
G005$-$040&5737& 4.02& $-$0.91&  0.11& $-$0.39&    31.4 & 54.7 & 35.8 & 18.8 & 17.4 & 14.2 & 18.2 & 26.0\\
G046$-$031&5907& 4.18& $-$0.81& $-$0.01& $-$0.40&    23.6 & 45.5 & 33.3 & 11.9 & 11.3 & 13.4 & 19.0 & 23.2\\
G088$-$040&5911& 4.14& $-$0.83&  0.17& $-$0.36&    30.4 & 59.3 & 43.9 & 17.9 & 19.1 & 14.1 & 19.1 & 24.6\\
G102$-$020&5310& 4.56& $-$1.09&  0.15& $-$0.39&    18.9 & 34.0 & 26.7 &  9.5 & 10.9 & 17.3 & 19.0 & 32.1\\
HD  3567&6041& 4.01& $-$1.20&  0.10& $-$0.49&    15.2 & 34.4 & 22.7 &  6.6 &  9.1 &  4.0 &  5.7 &  9.0\\
HD 17288&5700& 4.38& $-$0.88&  0.20& $-$0.36&    28.3 & 49.7 & 38.6 & 14.4 & 16.0 & 17.6 & 21.7 & 28.7\\
HD 17820&5750& 4.11& $-$0.69&  0.15& $-$0.30&    40.8 & 66.9 & 56.4 & 22.7 & 22.6 & 25.0 & 32.8 & 38.8\\
HD 22879&5774& 4.20& $-$0.86&  0.15& $-$0.40&    30.4 & 53.5 & 39.5 & 16.4 & 16.8 & 15.4 & 19.7 & 23.8\\
HD 24339&5810& 4.20& $-$0.67&  0.18& $-$0.35&    41.2 & 68.0 & 54.0 & 23.0 & 23.2 & 22.1 & 29.3 & 35.1\\
HD 25704&5765& 4.12& $-$0.91&  0.13& $-$0.32&    27.8 & 51.0 & 39.4 & 16.2 & 15.4 & 16.3 & 20.2 & 27.2\\
HD 76932&5849& 4.11& $-$0.88&  0.16& $-$0.36&    30.4 & 55.6 & 42.8 & 17.5 & 17.2 & 13.5 & 18.7 & 24.3\\
HD 83220&6470& 4.06& $-$0.49&  0.03& $-$0.23&         & 66.3 & 52.9 & 18.3 & 18.5  &     & 25.4 & 28.3\\
HD103723&6062& 4.33& $-$0.75&  0.07& $-$0.47&    23.6 & 48.7 & 36.6 & 13.6 & 12.4 & 11.3 & 14.8 & 18.4\\
HD105004&5832& 4.32& $-$0.78&  0.02& $-$0.34&    25.4 & 44.9 & 33.0 & 11.7 &      & 17.1 & 25.9 & 29.0\\
HD106038&5939& 4.23& $-$1.33&  0.17& $-$0.50&    13.1 & 29.5 & 18.3 &  6.1 &  4.8 &      &  4.2 &  8.0\\
HD113083A&5867& 4.35& $-$0.93&  0.16& $-$0.30&    22.0 & 42.2 &      & 12.1 & 10.8 & 14.7 & 18.1 & 24.8\\
HD113083B&5768& 4.34& $-$0.91&  0.11& $-$0.28&    22.4 & 43.7 & 29.3 & 11.9 & 12.1 & 17.9 & 22.7 & 29.6\\
HD113679&5595& 3.98& $-$0.71&  0.15& $-$0.32&    47.8 & 71.0 & 57.8 & 29.0 & 29.9 & 27.5 & 36.7 & 41.6\\
HD120559&5405& 4.40& $-$0.88&  0.13& $-$0.31&    25.1 & 45.8 & 35.6 & 13.5 & 16.1 & 26.2 & 34.3 & 43.9\\
HD121004&5622& 4.31& $-$0.71&  0.12& $-$0.37&    33.7 & 56.2 & 46.2 & 19.8 &      & 24.1 & 31.9 & 39.0\\
HD126681&5533& 4.28& $-$1.14&  0.07& $-$0.45&    15.9 & 31.2 & 24.8 &  7.3 &  9.8 &  9.8 & 15.4 & 18.8\\
HD241253&5830& 4.23& $-$1.10&  0.14& $-$0.45&    19.1 & 37.0 & 28.0 &  8.5 & 10.8 &  8.1 &  9.0 & 14.2\\
  W7547&6272& 4.03& $-$0.42&  0.07& $-$0.21&    49.8 & 78.1 & 66.0 & 32.5 & 27.2 & 27.7 & 35.3 & 40.2\\
\hline
\noalign{\smallskip}
\end{tabular}
\end{table*}
\end{document}